%% file: main_paper.tex
\documentclass{article}

\usepackage{microtype}
\usepackage{graphicx}
\usepackage{booktabs} 
\usepackage{multirow}
\usepackage{subfig}
\usepackage{comment}
\usepackage{xcolor}
\usepackage{amsmath}
\usepackage[flushleft]{threeparttable}
\usepackage{url}
\usepackage{hyperref}
\usepackage{amsfonts}
\usepackage[accepted]{mlsys2022}

\mlsystitlerunning{CryptoNite: Revealing the Pitfalls of End-to-End Private Inference at Scale}

\begin{document}

\twocolumn[
\mlsystitle{CryptoNite: Revealing the Pitfalls of End-to-End Private Inference at Scale}

% \mlsyssetsymbol{equal}{*}

\begin{mlsysauthorlist}
\mlsysauthor{Karthik Garimella}{nyu}
\mlsysauthor{Nandan Kumar Jha}{nyu}
\mlsysauthor{Zahra Ghodsi}{ucsd}
\mlsysauthor{Siddharth Garg}{nyu}
\mlsysauthor{Brandon Reagen}{nyu}

\end{mlsysauthorlist}

\mlsysaffiliation{nyu}{New York University}
\mlsysaffiliation{ucsd}{University of California, San Diego}

\mlsyscorrespondingauthor{Karthik Garimella}{kg2383@nyu.edu}

\mlsyskeywords{Machine Learning, MLSys}

\vskip 0.3in

\input{01abstract}
]



\printAffiliationsAndNotice{}  % leave blank if no need to mention equal contribution
% \printAffiliationsAndNotice{\mlsysEqualContribution} % otherwise use the standard text.

\input{03introduction}
\input{04preliminaries}

\input{05method}

\input{07experiments}

\input{09discussion}

\input{10related_work}

\input{11conclusion}

\input{ack}

\nocite{langley00}

\bibliography{references}
\bibliographystyle{mlsys2022}

%%%%%%%%%%%%%%%%%%%%%%%%%%%%%%%%%%%%%%%%%%%%%%%%%%%%%%%%%%%%%%%%%%%%%%%%%%%%%%%
%%%%%%%%%%%%%%%%%%%%%%%%%%%%%%%%%%%%%%%%%%%%%%%%%%%%%%%%%%%%%%%%%%%%%%%%%%%%%%%
% SUPPLEMENTAL CONTENT AS APPENDIX AFTER REFERENCES
%%%%%%%%%%%%%%%%%%%%%%%%%%%%%%%%%%%%%%%%%%%%%%%%%%%%%%%%%%%%%%%%%%%%%%%%%%%%%%%
%%%%%%%%%%%%%%%%%%%%%%%%%%%%%%%%%%%%%%%%%%%%%%%%%%%%%%%%%%%%%%%%%%%%%%%%%%%%%%%

%\appendix

%%%%%%%%%%%%%%%%%%%%%%%%%%%%%%%%%%%%%%%%%%%%%%%%%%%%%%%%%%%%%%%%%%%%%%%%%%%%%%%
%%%%%%%%%%%%%%%%%%%%%%%%%%%%%%%%%%%%%%%%%%%%%%%%%%%%%%%%%%%%%%%%%%%%%%%%%%%%%%%

\end{document}

%% file: 01abstract.tex
\begin{abstract}
The privacy concerns of providing deep learning inference as a service have underscored the need for private inference (PI) protocols that protect users' data and the service provider's model using cryptographic methods. Recently proposed PI protocols have achieved significant reductions in PI latency by moving the computationally heavy homomorphic encryption (HE) parts to an offline/pre-compute phase. Paired with recent optimizations that tailor networks for PI, these protocols have achieved performance levels that are tantalizingly close to being practical. In this paper, we conduct a rigorous end-to-end characterization of PI protocols and optimization techniques and find that the current understanding of PI performance is overly optimistic. Specifically, we find that offline storage costs of garbled circuits (GC), a key cryptographic protocol used in PI, on user/client devices are prohibitively high and force much of the expensive offline HE computation to the online phase, resulting in a 10-1000$\times$ increase to PI latency. We propose a modified PI protocol that significantly reduces client-side storage costs for a small increase in online latency. Evaluated end-to-end, the modified protocol outperforms current protocols by reducing the mean PI latency by $4\times$ for ResNet18 on TinyImageNet. We conclude with a discussion of several recently proposed PI optimizations in light of the findings and note many actually increase PI latency when evaluated from an end-to-end perspective. 

% Our code is available at \href{https://github.com/kvgarimella/cryptonite}{https://github.com/kvgarimella/cryptonite}.

%However, it incurs massive overhead in terms of latency, storage, and communication cost. Many crypto-protocols and design techniques have been proposed to {\em myopically} mitigate those overheads for a single PI assuming zero inference arrival rate  and overlooked the full-system effects; servicing the multiple inference requests coming at varying arrival rates in a real-world scenario. In this work, first, we study the full-system effect of state-of-the-art PI techniques and investigate their efficacy in a more realistic real-world scenario. Our study reveals the unexpected pain for end-to-end systems and the {\em key insight} is offline latency becomes the primary bottleneck for servicing the PI request beyond a certain arrival rate. Consequently, all online optimization benefits disappear after a certain unrealistically smaller arrival rate. To mitigate this, we propose the ``Client-Garbler'' PI method which can leverage the high storage capacity of the server (in a server-client setting) and substantially increases the serving rate of PI. Furthermore, we propose an arrival request rate-based network selection technique to forestall the queuing PI requests. We believe that our study would help in addressing the primary challenges and designing more efficient PI methods in a real-world scenario. 
\end{abstract}

%Many crypto methods for private inference (PI) produce impressive results, but ignore full-system effects; servicing multiple requests, storage and b/w costs as an example. Full-system evals reveal unexpected pain points. As example, protocols that use GC for ReLUs become storage limited when servicing multiple reqs; all offline optimization benefits disappear in this regime. We propose protocol switching, which is better suited from a full-system standpoint. Finally, we show that if reqs/sec vary during the course of a day, avg. error is minimized using model switching

%% file: 03introduction.tex
\section{Introduction}

Today, privacy continues to increase in relevance as users steadily demand more control over how and when their data is used and collected.
Addressing these concerns will likely have a profound effect on how the internet and many popular services, which heavily rely on precise user data for high-quality experiences, are implemented. One popular solution is to leverage privacy-preserving computation techniques that guarantee the privacy and security of user data during computations, thereby extending these guarantees from classical methods that only apply to communication. Privacy-preserving computation provides the best of both worlds for the user: privacy and security are both significantly improved and one still has access to online services integral to everyday life.
%methods that 
%\fixme{which enables computation directly on secure data}.
%With PEC, users are provided guarantees as to the privacy and security of their data that extends classic communication-only security and privacy to the execution of a function or program.

% Lots of work on private ML, describe space
% Systems/Stats/Crypto
% 2pc vs 3+PC, TPP
The scope of privacy-preserving solutions is vast.
Techniques vary in how they achieve security, whether by relying on system implementations (e.g., SGX \cite{intel_sgx}, Morpheous \cite{gallagher2019morpheus}), statistics (e.g. differential privacy \cite{dwork2006differential,abadi2016deep}), or cryptography (e.g. HE \cite{rivest1978data,gentry2009fully,lwe} and multi-party computation \cite{shamir1979share}).
Each comes with a tradeoff.
Relying on systems offers the best performance (e.g., Slalom \cite{tramer2018slalom}, DarKnight \cite{hashemi2020darknight}) but may concede security due to known vulnerabilities (e.g., Foreshadow \cite{van2018foreshadow}).
Statistical privacy is generally much stronger and has the benefit that it can be quantified;
for example using $\rho$ \cite{wood2018differential}.
However, it typically introduces approximation/noise to the computation, which can degrade inference accuracy, and is most applicable to answering questions in aggregate.
Therefore, its use in computing on individual user data is still unclear.
Cryptography-based solutions provide capabilities to compute exact inferences with very strong security, e.g., 128-bit, but come at the expense of high computational and communication/storage cost.
Homomorphic encryption (HE) and secure multi-party computation (MPC) are the two leading methods for cryptographically secure computation. 
%The challenge with HE and MPC, as we will detail in this paper, is that they incur extremely high overheads in nearly all aspects of computing, from run-time, to storage, and communication.
%There are also different variation in how protocols using these technologies are implemented, e.g., 2-party compute (2PC) or using a trusted third party (TTP).

% What we're looking at:
% 2pc crypto for inference
% exactly what that means
% excited because lots of promising recent work
 
%--- the problem of a user obtaining an inference result from a server-owned deep neural network (DNN) without the server learning the user's input and the user learning the server's DNN weights --- 
We explore private inference (PI)
using cryptographic primitives in a two-party compute (2PC) setting. 
In this setting, we assume that a client/user (the first party) has some data they would like to keep private while a server/service provider (the second party) performs inference on the user data using a deep neural network model, which they also want to keep private, on an encrypted representation of the data. We consider only 2PC protocols in this study because they require fewer trust assumptions than other protocols, e.g., trusted third party (TTP), and most closely mirror the plaintext client-server computational setting of today.
Our study is motivated by a growing body of work on PI that suggests PI run-times are approaching performance levels necessary for practical deployment. (In contrast, the overheads of private training are farther from practicality. Even so, our PI observations can inform developments in private training also.)

%This scenario has been adopted by many researchers, and a body of work now exists on using and optimizing cryptographic protocols for
%on PI. 
%We intend to look at training, but for now inference is enough as we cannot do training until we figure out inference.)
%cryptographic solutions as they offer the highest accuracy and security.

%PI has gained a lot of attention as deep learning models are dependent on large amounts of user data and are at the heart of many popular services.
%We elect to focus on inference as recent results suggest that PI performance 

% In this paper understand performance in end-to-end systems
% mimicking real use cases.
% Identify limitations under different, low, request arrival rates.

In this paper, we conduct the first end-to-end characterization of PI taking into account realistic system considerations\footnote{\textbf{This paper is the workshop version of this idea, see the full paper here} \cite{garimella22}}. 
Current 2PC PI protocols leverage a combination of cryptographic primitives, including HE, secret sharing (SS), garbled circuits (GC), and oblivious transfer (OT).
State-of-the-art (SOTA) solutions optimize for online runtime by
pre-computing SS constants using HE and preparing GCs in an offline phase, 
and then evaluating inference inputs using GCs and SS online.  

Instances of these protocols, including MiniONN \cite{liu2017oblivious}, Gazelle \cite{juvekar2018gazelle}, CrypTFlow2 \cite{rathee2020cryptflow2}, DELPHI \cite{mishra2020delphi}, CryptoNAS \cite{ghodsi2020cryptonas}, DeepReDuce \cite{jha2021deepreduce}, SAFENets \cite{lou2020safenet}, etc. (see Table \ref{tab:OptimizationComp} for a comprehensive list)
focus only on executing a \emph{single inference} at a time and mainly report improvements for \emph{online} computation/communication latency.
This leaves several open questions: 
(1) who pays for the offline costs, 
(2) what is the performance over multiple inference requests arriving at different rates, and 
(3) what is the role of storage cost?

%Of notable importance was the inclusion of inference request arrival rate and client-storage.
%Recent work on private inference,

%While significant strides have been made to improve performance

We find that the reported nearly-practical PI latency in recent work can be misleading as the pre-processing times of HE and storage requirements of GCs of each inference are so high that they can only truly be offline at extremely low arrival rates, e.g., 0.001 requests per second.
We observe that the standard ``Server-Garbler" protocols, as used in DELPHI~\cite{mishra2020delphi} and Gazelle~\cite{juvekar2018gazelle} CryptoNAS~\cite{ghodsi2020cryptonas} and DeepReDuce~\cite{jha2021deepreduce},
require more than 17KB to store GCs for \emph{each} ReLU in a DNN. 
Hence, for ResNet-18 on CIFAR-100, roughly 10GB (on TinyImageNet $\sim$40GB) of storage is needed for a \emph{single} inference (See Figure \ref{fig:SingleInference}c).
As a result, clients can only perform a (very) limited number of pre-computation before running out of storage capacity, limiting their ability to utilize the down-time between requests for pre-processing. This is a problem as without sufficient storage to buffer precomputations, the extreme runtime costs of HE must be incurred online, which can reduce online performance by 10-1000$\times$.
Hence, the offline pre-processing cost is quickly brought online as the system cannot keep up.

%Thus, clients can only store 1-2 pre-processed inferences, and 

%offline pre-processing is limited by the client device's storage capacity.

% Proposed optimizations
% protocol switching
Driven by the above analysis, we propose a protocol optimization to overcome the client-device storage limitations by reversing the roles of the client and server during the garbled circuit computation.
(We note that this has no impact on security, see Section 3.2 for details.)
By reversing the roles of the client and server we enable pre-computed GCs to be stored on the server, which has substantially more storage resources.
Using our optimized ``Client-Garbler" protocol we find that when assuming a server storage budget of 10 TBs, we can reduce the average PI latency by 4$\times$ when compared to the standard Server-Garbler protocol on TinyImageNet on ResNet-18.

% \fixme{reduce/improve} the average PI \fixme{latency/throughput?} by \fixme{X} compared to the standard Server-Garbler protocol.
%we can \fixme{double} the sustainable arrival rate of private inference requests.
%
% Removed for now, text makes us sound bad
%
% \fixme{I might just delete this: This optimization is not free.
% The tradeoff is that it requires moving one of the offline computations, specifically OT, to the online phase. 
% The result is a slightly increased online latency (20\%) for a \emph{single} inference, and in some cases, for \emph{very} low request arrival rates.
% \fixme{might make us sound a little bad.}
% Nonetheless, the protocols can be switched depending on the use case, and for most realistic use cases supporting high arrival rates is necessary.}

% Re-evaluate recent optimizations 
% (x2, cryptonas, deepreduce)
Finally, we expand our characterization of the optimized protocol to understand how recently proposed PI optimizations, which serve as examples of general design principles for efficient PI, impact performance in end-to-end systems.
We find that many papers focusing on ReLU optimization alone may not be sufficient and that some could even degrade performance (see Table \ref{tab:OptimizationComp} for a comprehensive list of prior work).
This is because while they can (significantly) reduce the storage pressure on the systems, they do not address extreme HE runtimes, which limit the rate that the offline precomputations can be done to replenish the supply.
Thus, we suggest future papers on PI optimization consider optimizing for \emph{both} ReLUs, to limit GC storage cost, and FLOPs, to mitigate high HE runtimes.

% We evaluate replacing ReLUs with x$^2$ (used in DELPHI and CryptoNets) to reduce ReLU/GC costs,
% CryptoNAS, which maximizes accuracy per ReLU by maximizing the FLOPs per ReLU in attempt to get the most expression of each,
% and finally DeepReDuce where ReLUs are selectively pruned from the network.
% \fixme{Brief discussion of findings.}

% We conclude that..
We conclude that recent low-latency MPC protocols for PI (including optimizations) are in the end performance limited by HE.
When considering arrival rates, or more than one inference in isolation, and practical storage limitations the offline cost quickly overwhelms client/server systems and cause the offline cost to be brought online.
HE costs are extreme, e.g., it can take roughly 20 minutes to perform inference on TinyImageNet for ResNet-18, and suggests the need for more work in systems support for HE or looking into secure alternatives for computing secrets.

This paper makes the following contributions:
\begin{itemize}
\item We conduct the first systems-level study for state-of-the-art PI techniques. 
Our experiments using different inference request arrival rates reveal client storage limitations and offline homomorphic encryption times limit performance, even when using optimized MPC protocols for PI. 
\item To overcome the client-storage limitations, we propose the ``Client-Garbler'' PI protocol.
By switching the roles of parties in existing PI protocols (e.g., as used in DELPHI and Gazelle) we are able to reduce the mean PI latency by $4\times$ for ResNet-18 on TinyImageNet.
\item Finally, we evaluate recently proposed optimizations for private inference performance, e.g., replacing ReLU with $x^2$~\cite{gilad2016cryptonets, mishra2020delphi} and ReLU dropping~\cite{jha2021deepreduce}.
We find that when considering end-to-end system-level effects many optimizations offer limited benefit.
Our analysis motivates the need for more research into reducing both ReLU and FLOP costs in DNN models.
\end{itemize}

%The precomputation necessary for \fixme{actually not clear to me}
%as a shield for HE, quickly broken.

%% file: 04preliminaries.tex
\section{Preliminaries} \label{sec:prelims}

In this section, we provide a brief summary of the cryptographic building blocks used for privacy-preserving inference and show how they can be combined to construct an end-to-end protocol.

\subsection{Cryptographic Primitives}\label{sec:crypto}

%\note{from circa rewrite }\textbf{Finite Fields.} The cryptographic primitives described subsequently operate on values in a finite field of integer modulo a prime $p$, $\mathbb{F}_{p}$, i.e., the set $\{0,1,\ldots,p-1\}$. In practice, positive values will be represented with integers in range $[0,\frac{p-1}{2})$, and negative values will be integers in range $[\frac{p-1}{2},p)$. %don't need

\textbf{Additive Secret Sharing.} Additive secret sharing (SS) is a protocol that allows a value $x$ to be split amongst two (or more) parties such that neither party learns anything about $x$ from its share while the two parties can reconstruct $x$ if they work together. 

Specifically, the shares of a value $x$ are %can be created for two parties by randomly sampling a value $r$ and setting shares as 
$\langle x\rangle_1 = r$ and $\langle x\rangle_2 = x-r$, where $r$ is a random value and the arithmetic operations are module a prime $p$. 
Since $x=\langle x\rangle_1+\langle x\rangle_2$, the value $x$ can be recovered if the two parties add their shares together. However, since $r$ is random, neither party has enough information to reconstruct $x$ by itself.

Given shares of two values $x$ and $y$, shares of $x+y$ can be computed by as follows:  $\langle x + y\rangle_1 =  \langle x \rangle_1 + \langle y\rangle_1$, and $\langle x + y\rangle_2 =  \langle x \rangle_2 + \langle y\rangle_2$. That is, the shares of a sum of two values equal the sum of the shares. 
%Performing additions over two shared values is straightforward in this scheme, each party simply adds their respective shares to obtain an additive sharing of the result.

%\note{from circa rewrite }\textbf{Beaver Multiplication Triples.} This protocol~\cite{beaver1995precomputing} is used to perform multiplications over two secret shared values. A set of multiplication triples are generated offline from random values $a$ and $b$, such that the first party receives $\langle a\rangle_1$, $\langle b\rangle_1$, $\langle ab\rangle_1$, and the second party receives $\langle a\rangle_2$, $\langle b\rangle_2$, $\langle ab\rangle_2$. 
%In the online phase $x$ and $y$ are secret shared among parties such that the first party holds $\langle x\rangle_1$, $\langle y\rangle_1$ and the second party holds $\langle x\rangle_2$, $\langle y\rangle_2$. To perform multiplication they consume a set of triples generated offline and at the end of the protocol the first party obtains $\langle xy\rangle_1$ and the second party obtains $\langle xy\rangle_2$. %don't need

\textbf{Homomorphic Encryption.} HE is a cryptographic protocol that allows a client to encrypt its data in a manner that allows a server to compute a function on the client's data without ever decrypting the data. 
HE involves three steps. First, the client encrypts data $x$ using encryption function $E$ and its public key $k_{pub}$, i.e., $c = E(x,k_{pub})$ (or just $c = E(x)$ for short).
Then, given two cipher-texts 
$c_1 = E(x_1)$ and $c_2 = E(x_2)$, the server homomorphically computes $x_1 \odot x_2$ using a function $\star$ that operates directly on ciphertexts, i.e., $c'=c_1 \star c_2$.  The client can then decrypt $c'$ using decryption function $D$ and secret key $k_{sec}$ to obtain $x_1 \odot x_2 = D(c',k_{sec})$ (or just $D(c')$ for short). We will also write $c_1 \star c_2$ as $HE(x_1 \odot x_2)$ for clarity.

\textbf{Oblivious Transfer.} OT~\cite{rabin2005exchange} is a fundamental building block in MPC and the foundation for many constructions. In 1-out-2 OT there are two parties: the sender and the receiver. The sender has two input values $x_0, x_1$, and the receiver takes as input only a selection bit $r$. The OT protocol allows the receiver to obtain one of the sender's inputs corresponding to their selection bit ($x_r$). The sender will not learn any information, and the receiver will not learn the other input value ($x_{1-r}$). Although OT protocols require expensive public-key cryptography, it is possible to use a small number of \emph{base OTs} and extend them to achieve a larger number of OTs using cheaper symmetric key cryptography operations~\cite{ishai2003extending}. These are referred to as \emph{OT extensions}. 
%based OTs between the sender and receiver with their roles reversed where the receiver's inputs to the base OT are (typically 128) pairs of randomly generated vectors (or seeds), and the sender's input is a random selection vector. These random vectors are then used during OT extension to transfer values obliviously from the sender using simple hash functions. \zg{ot ext is hard to explain, I wonder if we should just describe at a higher level}

\textbf{Garbled Circuits.} The GC protocol was first introduced by Yao~\cite{yao1986generate} and it enables two parties to compute a {Boolean} function on their private inputs without revealing their inputs to each other. The function is first represented as a Boolean circuit, and one party (the garbler) assigns two random labels (keys) to each wire of the circuit, corresponding to values $0$ and $1$. 
For each gate, the garbler then generates an encrypted truth table that maps the output labels to the gate's input labels.
%encrypts each output key with the corresponding input keys, obtaining garbled tables. 
The garbler sends the generated garbled circuit to the other party (the evaluator), along with the labels corresponding to its inputs. 
The evaluator then uses the OT protocol to obtain the labels corresponding to its inputs 
without the garbler learning the input values.
At this point, the evaluator can execute the circuit gate by gate, without learning intermediate values. 
Finally, the evaluator shares the output labels with the garbler who maps them to plaintext values.

\input{plots/delphi_protocol}

\subsection{Protocol for Private Inference}\label{sec:privateinf}
We begin  by reviewing  a class of PI protocols that
use HE and SS
for linear layers, and GCs for ReLU layers~\cite{liu2017oblivious, juvekar2018gazelle, mishra2020delphi}. These protocols have achieved SOTA results in the 2PC setting.
In this setting, we assume a client and a server who wish to obtain the server's model prediction on the client's input while preserving the privacy of both the model and the input. 
Following prior work, we assume both parties are honest-but-curious, i.e., they follow the protocol faithfully but try to extract information during protocol execution.

We provide a brief description of the DELPHI protocol below, with a depiction in Figure~\ref{fig:protocol}(a). 
The protocol consists of an offline phase, which only depends on the network architecture and parameters, and an online phase, which is performed after the client's input is available. 
We represent model parameters at layer $i$ with $\mathbf{W_i}$ (omitting biases in the description for simplicity). The layers after each linear and ReLU operation are represented by $\mathbf{x_i}$ and $\mathbf{y_i}$, respectively.
At a high level, the protocol allows the client and the server to hold additive shares of each layer value during inference evaluation.
We assume DELPHI as a baseline (Server-Garbler) protocol in this paper.

\textbf{Offline Phase.} 
During the offline phase, the client generates the encryption keys and sends them to the server. The client and the server then sample random vectors per each linear layer 
%from $\mathbf{F}^n_p$
from a finite ring,
denoted by $\mathbf{r_i}$ and $\mathbf{s_i}$ respectively. The client encrypts its random vectors and sends them to the server. After obtaining $\textrm{E}(\mathbf{r_i})$, the server computes $\textrm{E}(\mathbf{W_i}.\mathbf{r_i}-\mathbf{s_i})$ homomorphically and sends it back to the client. The client decrypts the ciphertext received from the server and stores the plaintext values.
For each Nonlinear ReLU operation, the server constructs and garbles a circuit that implements $\textrm{ReLU}(\mathbf{x_i})-\mathbf{r_{i+1}}$, where $\mathbf{r_{i+1}}$ is the next layer randomness (provided by the client). 
The server sends the garbled circuits to the client and the two parties engage in the base OT protocol after which the client obtains labels corresponding to its inputs through the OT extension protocol.
At the end of the offline phase, the server stores its random vectors $\mathbf{s_i}$ and the garbled circuit input and output encodings, and the client stores its random vectors $\mathbf{r_i}$, garbled circuits, and the labels corresponding to its inputs. The communication and storage requirements for ResNet-32 inference on a single input from CIFAR-10 are shown in Figure~\ref{fig:protocol}(a). 
As depicted here, the largest storage requirement is $5.3$ GB on the client due to garbled circuits.

\textbf{Online Phase.} 
The client's input $\mathbf{y_1}$ is available in the online phase. The client computes $\mathbf{y_1}-\mathbf{r_1}$ and sends it to the server. At the beginning of $i$th linear layer, the server holds
$\mathbf{y_i}-\mathbf{r_i}$ and the client holds $\mathbf{r_i}$.
The server computes its share of layer output as $\langle \mathbf{x_i} \rangle_{s} = \mathbf{W_i}(\mathbf{y_i}-\mathbf{r_i})-\mathbf{s_i}$. 
The client holds $\langle \mathbf{x_i} \rangle_{c} = \mathbf{W_i}.\mathbf{r_i}+\mathbf{s_i}$ from the offline phase, and the client and the server hold additive shares of $\mathbf{x_i}=\mathbf{W_i}.\mathbf{y_i}$. To evaluate the ReLU layer, the server obtains the labels corresponding to its share of ReLU input $\langle \mathbf{x_i} \rangle_{s}$ and sends them to the client. The client now holds labels for the server's input, as well as labels for its inputs $\langle \mathbf{x_i} \rangle_{c}$ and $\mathbf{r_{i+1}}$, which were obtained during the offline phase along with the garbled circuits. The client evaluates the circuit and sends the output labels to the server who is then able to obtain
$(\mathbf{y_{i+1}}-\mathbf{r_{i+1}})$. At this point, the client and server hold additive shares of $\mathbf{y_{i+1}}$ and will similarly evaluate subsequent layers.

%% file: plots/delphi_protocol.tex
\begin{figure*} [t]
  \centering
  \subfloat[Server-Garbler]{\includegraphics[scale=0.25]{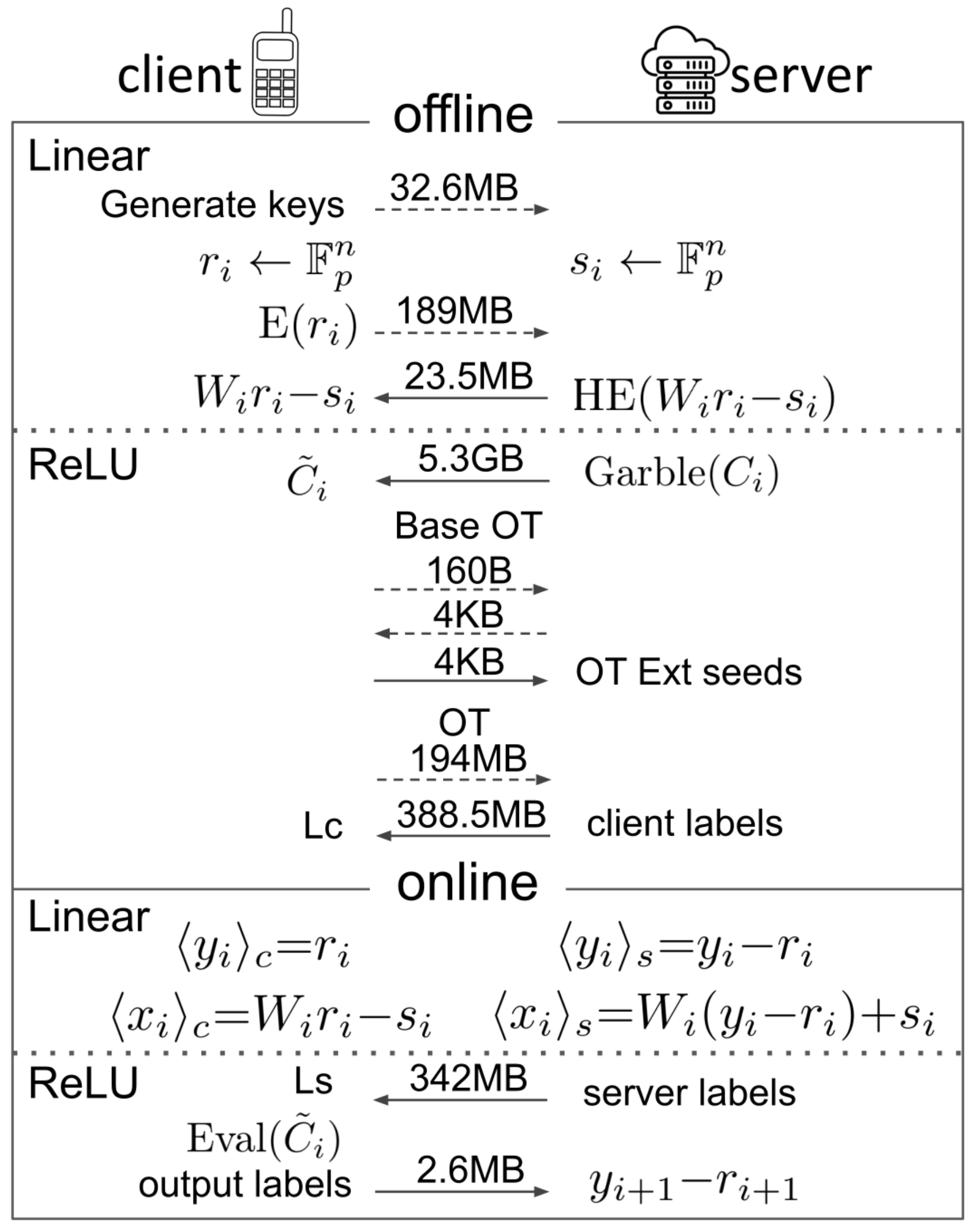}}
  \hspace{1cm}
   \subfloat[Client-Garbler]{\includegraphics[scale=0.25]{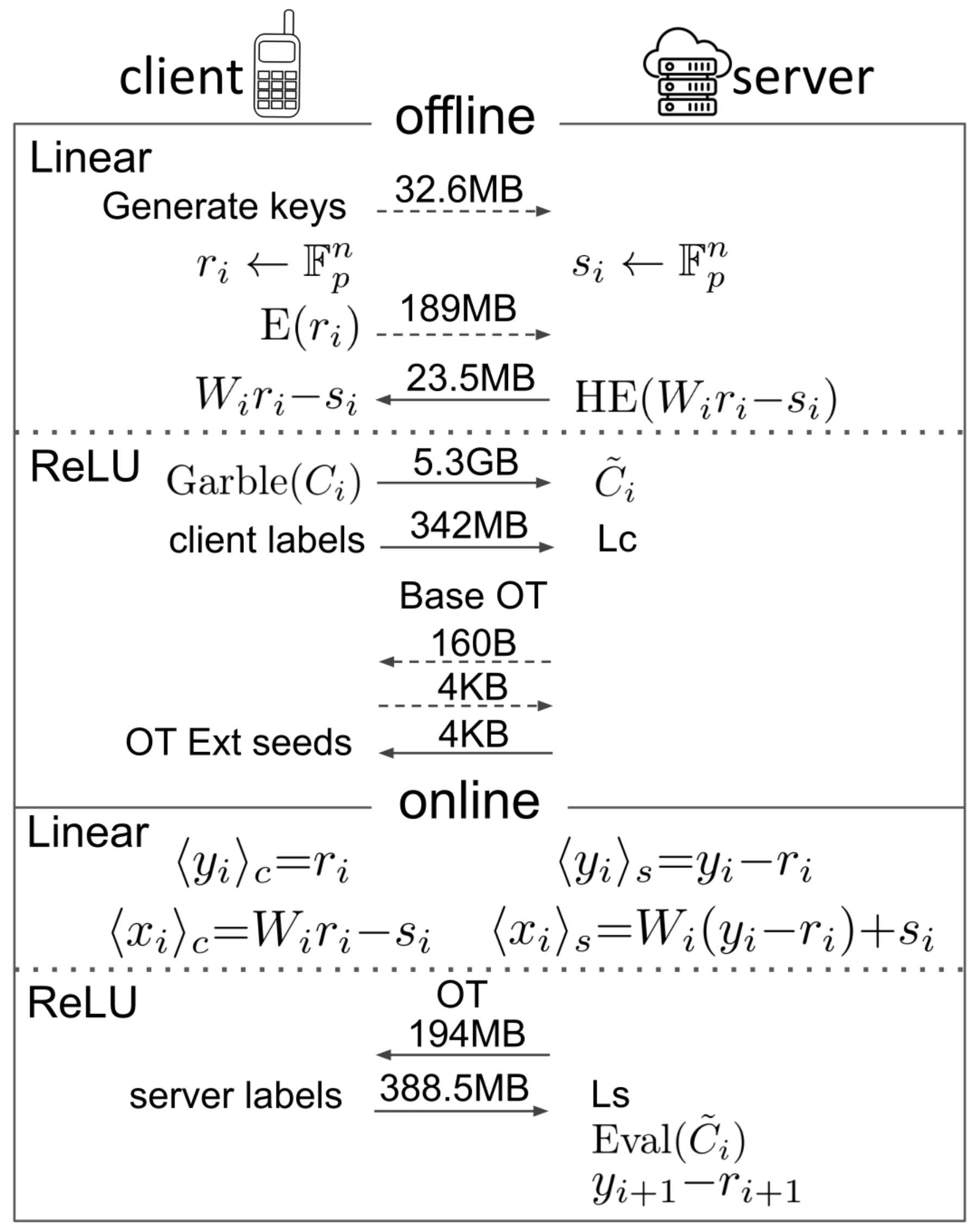}} 
  \caption{(a) The baseline ``Server-Garbler'' protocol used in prior work. Solid arrows in the offline phase of the protocol represent data that are stored by the recipient, while data communicated via dashed arrows can be discarded. (b) Proposed ``Client-Garbler'' protocol. Observe the significant reduction in offline storage on the client-side.  }%\vspace{-10pt}
  \label{fig:protocol}
\end{figure*}

%% file: 05method.tex
\section{Method} \label{sec:method}

In this section we first motivate our study by noting observations about the assumed baseline Server-Garbler protocol from existing work
and how the storage and total performance costs limit its practical deployment. 
Next, we describe the proposed optimized Client-Garbler protocol and its benefits over the baseline solution.

\input{plots/pie_chart_breakdown}

\subsection{Motivational Observations}
Figure~\ref{fig:SingleInference}(a) and (b) respectively show the breakdown of computation and communication costs of the baseline PI protocol evaluated on a \emph{single} inference, which is typical of prior work. 
The computation costs are dominated by HE, while the communication costs are dominated by GC. 
The key observation in DELPHI~\cite{juvekar2018gazelle} was that HE computations involve only the server's DNN model weights but not the client's inputs. Therefore, assuming that the same model is used for multiple inferences, the computation costs of HE can be
moved to input independent pre-compute/offline phase. Given that GC's communication costs are always incurred offline, this means the online latency for a single inference can be reduced to roughly online GC computation costs and online OT communication costs, both of which are small. Table~\ref{tab:LatencyCost} shows that total online latency (including computation and communication latency) is more than $10\times$ smaller than the offline latency. 

Thus, as long as the client has sufficient downtime to perform pre-computations, it can achieve relatively fast inference results when requests arrive. 
For instance, ResNet-32 on CIFAR-100 has an online latency of 9.4s; while still far from real-time, a sub-10s latency is an encouraging result considering the strong security guarantee that PI provides. 

However, the storage costs are shown in Figure~\ref{fig:SingleInference}(c) paint a  different picture. The ResNet-32 model on CIFAR-100 incurs a $5.3$GB storage cost, most of which is to store the garbled circuit on the client device assuming a standard server garbler protocol.
(We stress here that although the garbled circuit is pre-computed, it cannot be used or amortized over multiple inferences. Each pre-computed garbled circuit can only be used for a single online inference and a new circuit must be pre-computed for the next inference.) Consider a client device with 8GB storage dedicated for PI.
Such a device would only have the capacity to store a \emph{single} pre-computation. 
This means that the client device would not have the opportunity to leverage any idle time between requests, due to either low request rates or variation in arrival times, to pre-compute and buffer computations, as there is simply no space to store them.
Thus, as the request rate increases, large parts of the offline costs will be incurred online.

The picture is even bleaker for the TinyImageNet and ImageNet datasets. Depending on the network, the former requires between 14.7GB-38.9GB of largely client-side storage, a stretch for even high-end mobile phones, while the latter's 180GB (or greater) storage costs are currently infeasible on most modern devices. 
Thus, private inference on ImageNet using existing Server-Garbler protocols cannot make use of pre-compute, moving the already large offline costs fully online.

\subsection{Client-Garbler Protocol}
\label{sec:cg3}
In this section, we introduce our \emph{Client-Garbler} protocol, which modifies the protocol described in Section~\ref{sec:privateinf} to alleviate the high storage requirements on the client. As discussed previously, this storage cost is largely due to the size of garbled circuits. We propose to \emph{switch} the roles of the garbler and evaluator between the server and the client. Here the client plays the role of the garbler and the server plays the role of the evaluator. 
Hence, in our Client-Garbler protocol,
the garbled circuits will be stored on the server, which we assume to have access to high-capacity storage. Our proposed Client-Garbler protocol is depicted in Figure~\ref{fig:protocol}(b). The linear layers (in offline and online phases) are processed as before. 

During the offline phase for ReLU layers, the client garbles the circuits and sends them to the server along with the labels corresponding to its inputs (known in the offline phase). 
However, in this new protocol, the server can no longer acquire labels corresponding to its own inputs from the client in the offline phase since they depend on $x$, which is not known offline. The server must therefore wait to obtain these from the client \emph{online} using OT while base OTs can however be performed offline. 
For ResNet-32 on CIFAR-100, the client's offline storage costs in the client garbler protocol are only $23.5$ MB while $5.3$ GB is stored on the server for a ResNet-32 (Figure~\ref{fig:protocol}(b)).

During the online phase, the server computes the linear layer and obtains its share of the output $\langle \mathbf{x_i} \rangle_{s}$.
The client and the server then engage in OT extension protocol and the server obtains the labels corresponding to its input. The server is now able to evaluate the GC and obtain the output, its share of the subsequent linear layer. The introduction of OT in the online phase can slightly increase online costs compared to Server-Garbler protocols.
However, in return, we obtain massive reductions in client storage,
and in end-to-end system evaluations, this results in a net performance benefit for the proposed Client-Garbler protocol. 
Still, when PI is evaluated from only the lens of a single inference, the server garbler protocol is slightly better. 
We note that both protocols have the same security because the security guarantees of GC do not depend on which party is the garbler.

%% file: plots/pie_chart_breakdown.tex
\begin{figure*} [t]
\centering
\subfloat[ResNet-18 latency breakdown]{\includegraphics[scale=0.32]{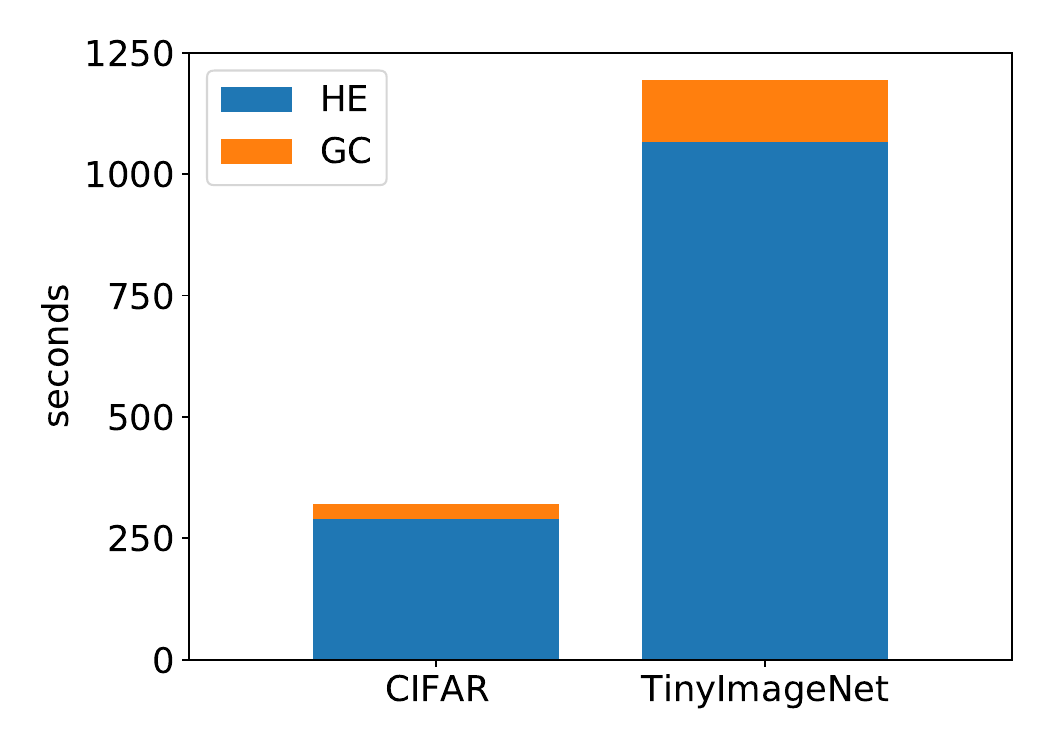}}
\subfloat[ResNet-18 communication breakdown]{\includegraphics[scale=0.32]{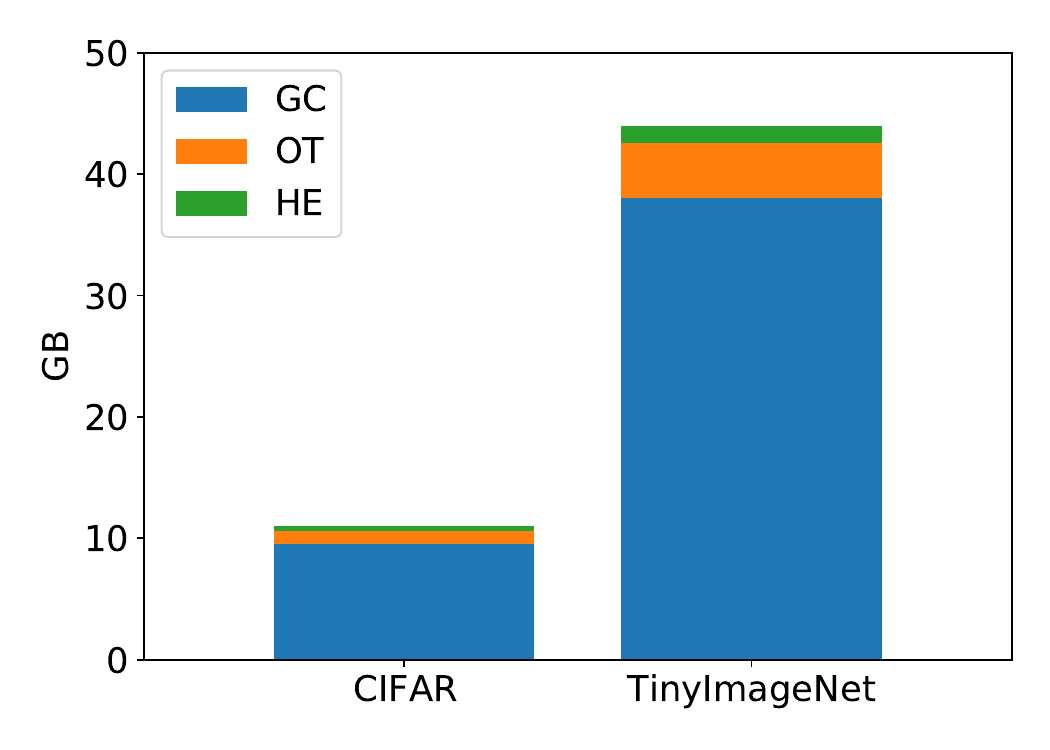}} 
\subfloat[Garbled Circuit Storage Requirement]{\includegraphics[scale=0.32]{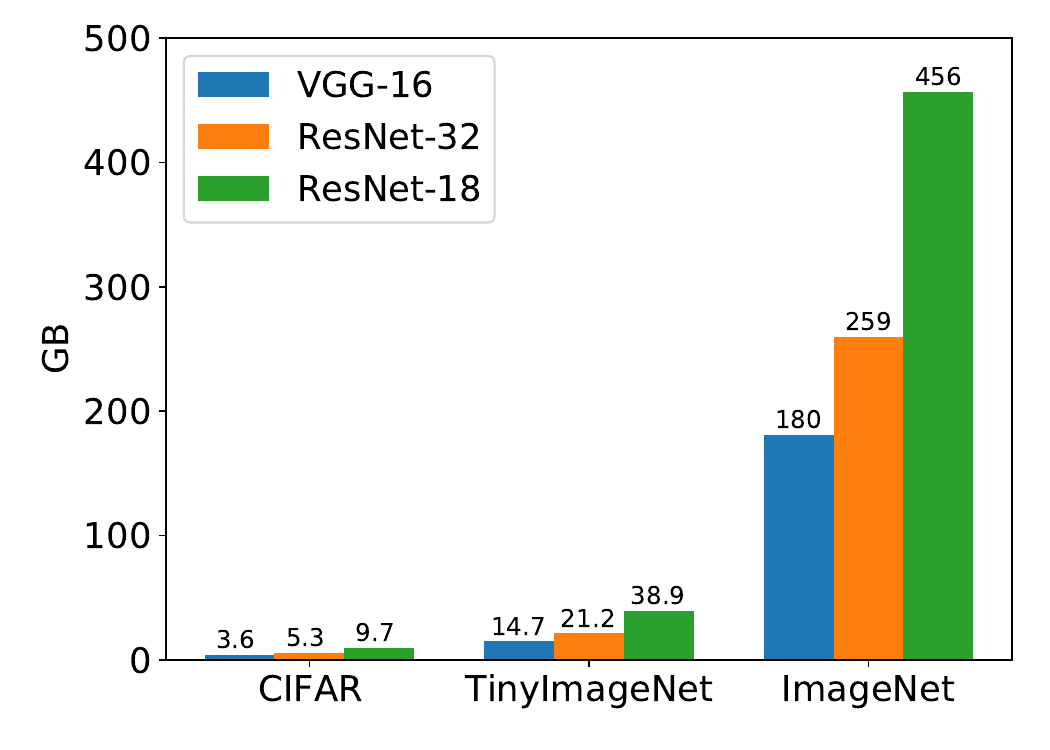}} 
\caption{Latency and communication breakdown (Fig. a and Fig. b) for end-to-end \textit{single image} encrypted inference on CIFAR-100 and TinyImageNet, and storage overhead (Fig. c) for the same on CIFAR-100, TinyImageNet, and ImageNet. Homomorphic evaluation of the linear layers dominates the compute latency, while Garbled Circuits dominate the communication between the server and client.}
\label{fig:SingleInference}
\end{figure*}

%% file: 07experiments.tex
\section{Experimental Results} \label{sec:results}

\input{tables/NetArch}

{\bf Networks and Datasets:}
We performed experiments with ResNet models \cite{he2016deep},  ResNet-32 and ResNet-18, and VGG-16 \cite{vgg} and we made the same architectural modifications to these networks as made in \cite{jha2021deepreduce} and \cite{ghodsi2021circa}. In particular, we remove the downsampling in earlier layers and replaced all the max-pooling operations with average-pooling as the latter is a linear operation and cheaper for PI. The layer, parameter, FLOPs, and ReLU counts for these networks are described in Table \ref{tab:NetArch}. We perform our experiments of CIFAR-100 \cite{cifar} and TinyImageNet \cite{yao2015tiny} dataset with 100 and 200 class labels, respectively.  CIFAR-100 has images of spatial resolution $32\times32$ while TinyImageNet is more complex and has images of resolution  $64\times64$.

{\bf Experimental Setup:}
We develop a private inference simulator using the Simpy library to explore the full system effects of PI at scale \cite{simpy}. We simulate both the conventional Server-Garbler and our proposed Client-Garbler protocol with a default bandwidth of 100 MB/sec. In our evaluations, we sweep client-side storage capacity from $8$GB to $256$GB and assume a $10$TB storage on the server-side. 
We assume Poisson arrivals of inference requests at a given rate. 

For a specified protocol, network, and dataset, we simulate 24 hours of inference requests for a fixed arrival rate (assuming Poisson arrivals) and average inference latency over 100 simulation runs. Two processes are simulated concurrently: 1) the client and server check if enough storage is available to perform an offline phase in order to obtain precompute and Garbled Circuits and 2) the client and server engage in the online phase of either protocol if a private inference has been requested. 

We model a single-client and single-server system where inference requests are queued and served in a First-In-First-Out manner. When an inference request moves out of the queue, the simulator checks the storage capacity for previously computed Garbled Circuits. If enough Garbled Circuits are available (for a single inference), the client and server begin executing the online phase of the configured protocol. If Garbled Circuits are not available, the client must wait for an offline phase to complete before performing the online phase. 

We computed the latency, storage, and communication costs for the online and offline phase of both protocols using the DELPHI codebase. For all networks, we averaged these costs over 10 and 5 runs for CIFAR100- and TinyImageNet-like inputs, respectively. Finally, we input these costs into the simulator for both the offline and online phases. 

\input{plots/mean_latency_comparison}
\input{plots/offline_bottleneck}

\input{tables/LatencyCost}

\subsection{Bottleneck 1 - GC Storage}
% {\bf Bottleneck 1 - GC Storage:}
Figure~\ref{fig:CompCase2AndCase1} compares the Server-Garbler and Client-Garbler protocol for the aforementioned networks and datasets. In particular, for the Server-Garbler protocol, we sweep both arrival rates and storage capacities. Client-side storage capacities range from 8 GB to 64 GB for CIFAR-100. For the proposed Client-Garbler protocol, we only simulate the smallest 8GB client-side capacity because this suffices to hold more than 300 pre-computes. The results are plotted in Figure~\ref{fig:CompCase2AndCase1}a-c.

For all arrival rates, the proposed Client-Garbler protocol exhibits a lower mean inference latency when compared to the Server-Garbler protocol with 8 GB capacity (blue). This is because, with only 8 GB of storage for the Server-Garbler protocol, the client can only store at most one precompute for both ResNet-32 and VGG-16. We note that for ResNet-18, the client is unable to store even a single precompute with 8GB capacity as a single inference requires more than 9 GB of garbled circuits. Hence, for this network, we only plot results for 16 GB client-side storage and above. 

%Thus even at a low arrival rate, inference requests in the queue will incur some proportion of the offline phase. However, the Client-Garbler protocol can leverage the high storage capacity of the server. 

For all networks, at 64 GB of client-side storage capacity (red), the mean inference latency for the Server-Garbler protocol is slightly lower than or similar to the Client-Garbler protocol, since enough capacity is now available to store GC precomputes without saturating the queue. That is, even for CIFAR-100, the Server-Garbler needs up to 64 GB of storage to ``catch up" to the Client-Garbler protocol which uses less than 8 GB storage. Every simulation begins with an empty storage capacity. 
%In all storage scenarios, As the arrival rate increases, the mean latency for the Server-Garbler protocol increases rapidly as waiting time in the queue increases. The mean latency of the Client-Garbler protocol remains low (due to high server-side storage capacity) before exhibiting a significant increase in mean latency.

For the TinyImageNet dataset (see Figure~\ref{fig:CompCase2AndCase1}d-f), we perform experiments with client-side storage capacity from 64 to 256 GB as the size of Garbled Circuits for a single inference ranges from 15 to 40 GB. Similar characteristics as to the CIFAR-100 experiments are observed for each network.  We note a $4\times$ reduction in mean PI latency at the arrival rate of 0.004 request/sec for ResNet-18 on TinyImageNet dataset.

\subsection{Bottleneck 2 - Offline HE Latency}
% {\bf :}
The Client-Garbler protocol enjoys the benefits of high capacity server-side storage for Garbled Circuits and precomputes, which can be observed at moderate arrival rates. However, at even higher arrival rates, the mean latency for the Client-Garbler protocol reaches on the order of hours for CIFAR-100. Figure~\ref{fig:SgCgWaitingTime} shows the breakdown of mean latency for a single inference request for three arrival rates for ResNet-18 on CIFAR-100. As the arrival rate increases, more cycles are spent waiting for offline precomputes to become available and previous inference requests to complete. As shown in Figure~\ref{fig:SingleInference}a, HE operations account for over 90\% of latency during the offline phase. This issue is further aggravated by networks with higher FLOPs on complex datasets such as ResNet-18 on TinyImageNet. Thus, the HE overhead cannot be mitigated at higher arrival rates and effectively becomes an online computation.

%% file: tables/NetArch.tex
\begin{table} [t]
\caption{Network architecture and number of parameters, FLOPs, ReLUs, and accuracy (Acc(C100)) on CIFAR-100 . Number of layers are in the order of (\#Conv, \#ReLU, \#Avgpool, \#FC) layers.}
\vspace{.5em}
\label{tab:NetArch}
\begin{threeparttable}
\centering 
\resizebox{0.49\textwidth}{!}{
\begin{tabular}{lccccc} \toprule
Model & \#Layers & \#Params & \#FLOPs & \#ReLUs &Acc(C100) \\ \toprule
ResNet-32 & (31, 31, 1, 1) &0.5M & 68.9M & 303.1K & 67.03\%\\
VGG-16 &  (13, 15, 5, 3) & 34M & 332.5M & 284.7K & 73.45\%\\  
ResNet-18 & (17\tnote{*}, 17, 1, 1) & 11M & 555.5M & 557.1K & 74.20\% \\ \bottomrule
\end{tabular} }
\begin{tablenotes} \tiny
            \item[*] There are three additional $1\times1$ Conv layers in the skip connections which are not include here.
            \end{tablenotes}
     \end{threeparttable}
\end{table}

%% file: plots/mean_latency_comparison.tex
\begin{figure*} [t]
\centering
\subfloat[CIFAR-100, ResNet-32]{\includegraphics[scale=0.25]{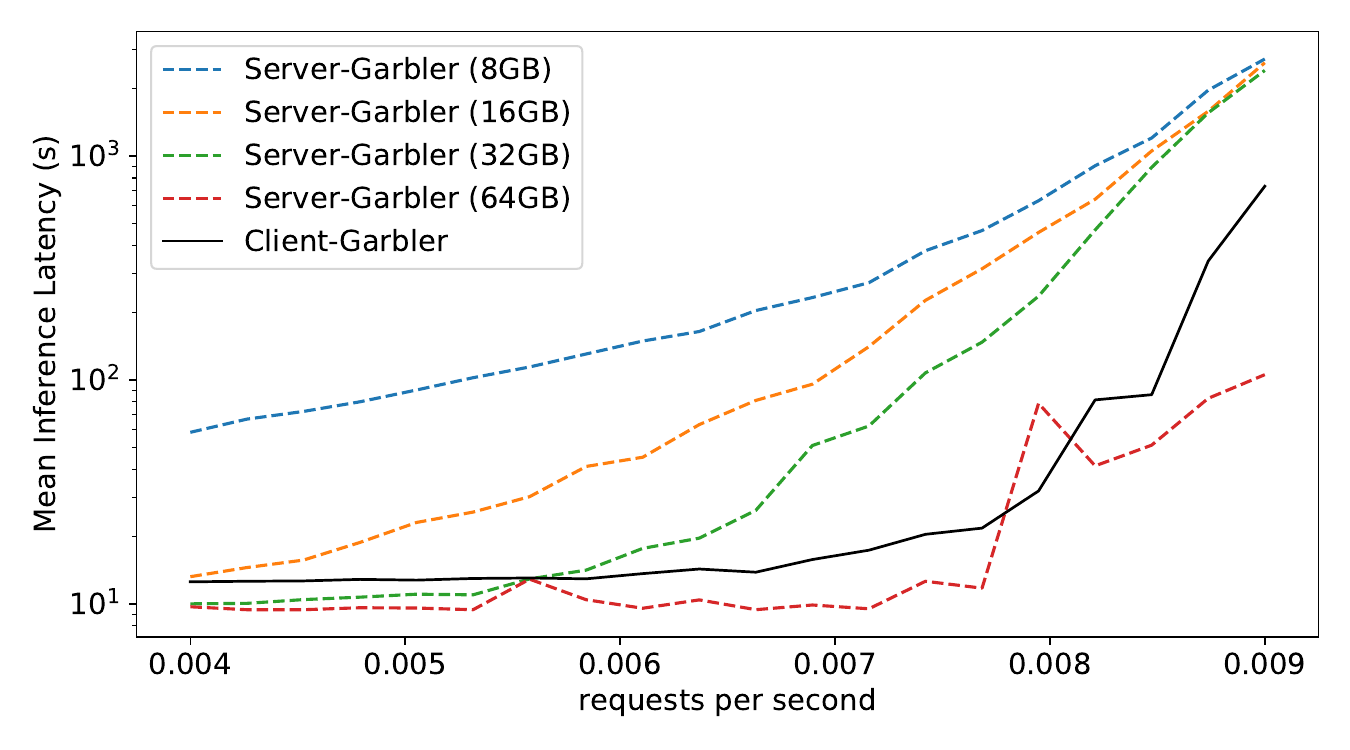}}
\subfloat[CIFAR-100, VGG-16]{\includegraphics[scale=0.25]{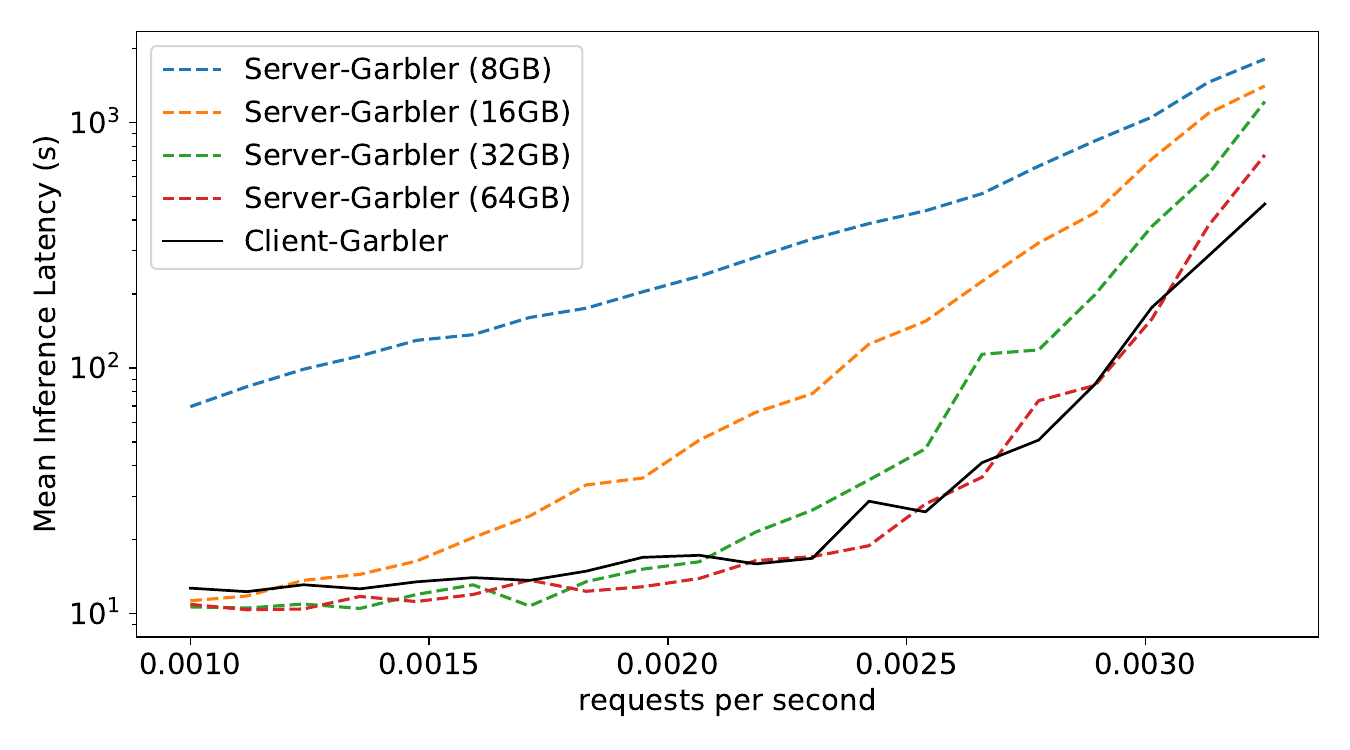}} 
\subfloat[CIFAR-100, ResNet-18]{\includegraphics[scale=0.25]{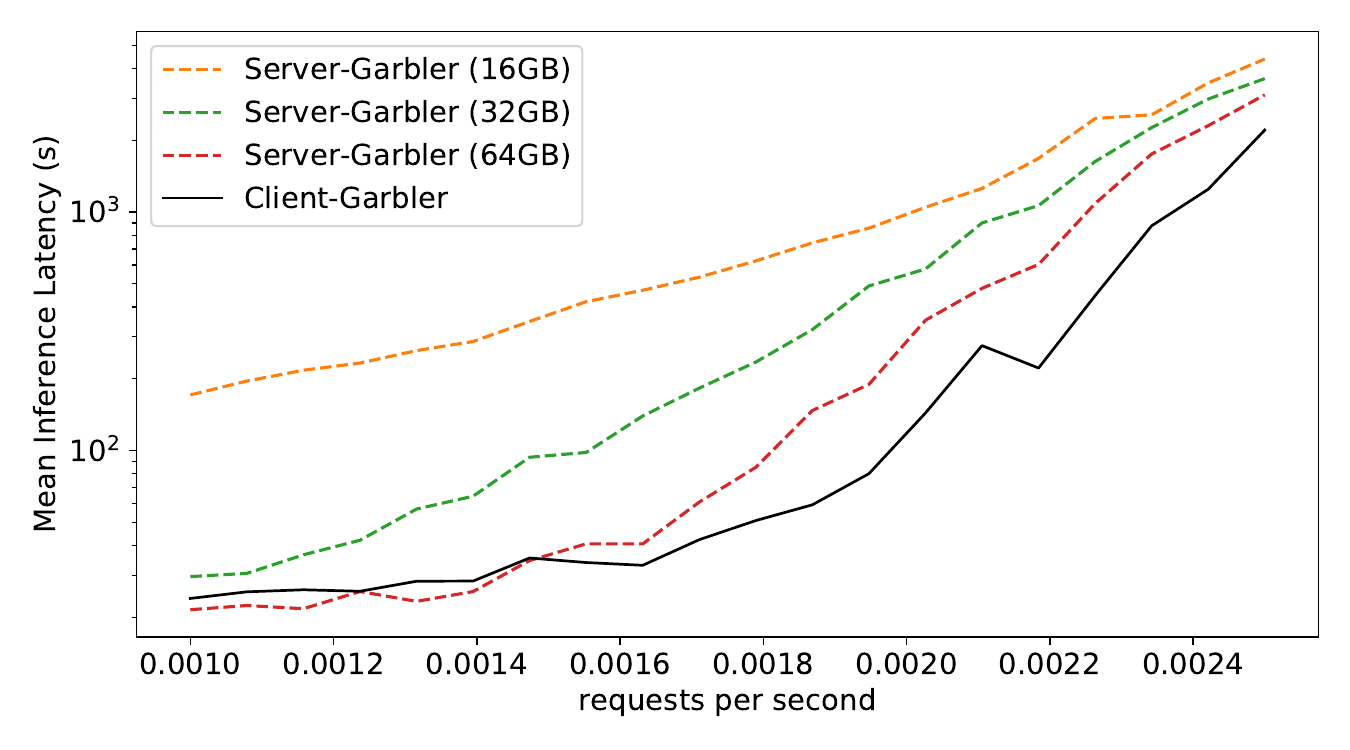}}
\\
\subfloat[TinyImageNet, ResNet-32]{\includegraphics[scale=0.25]{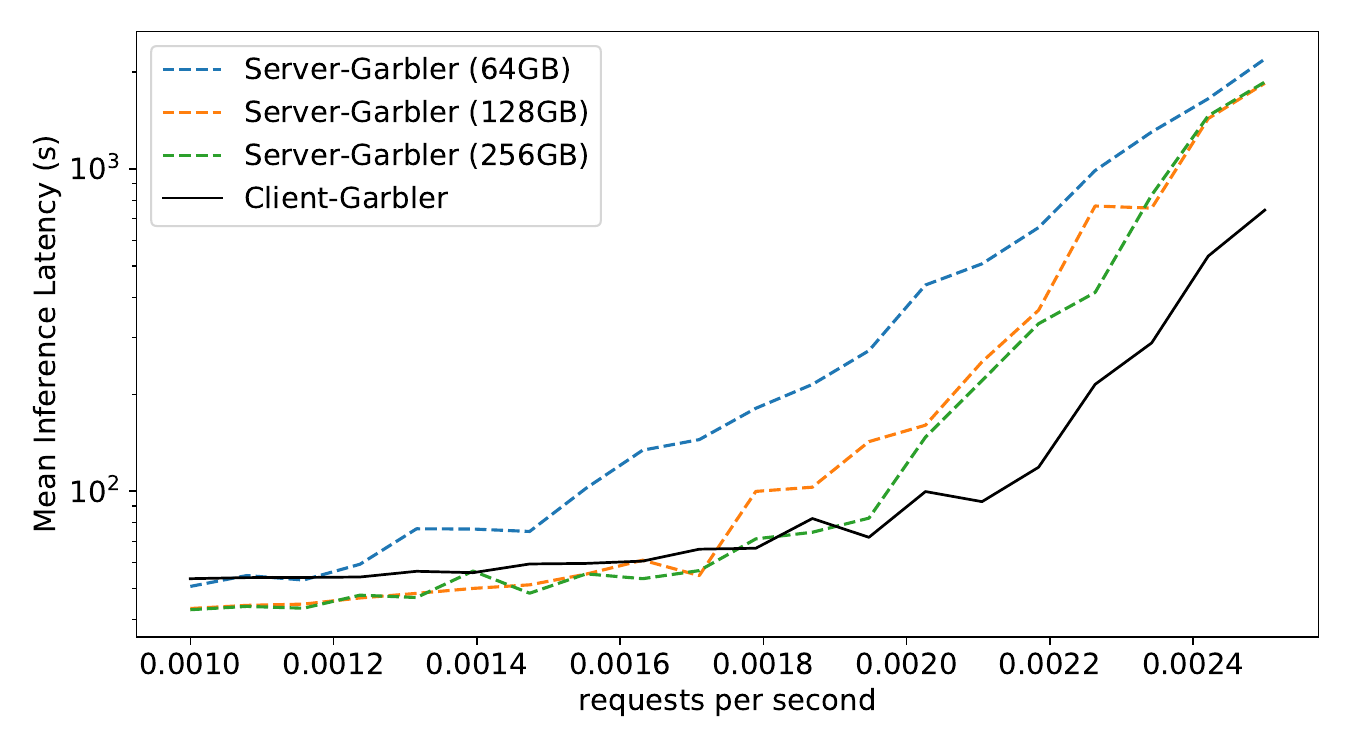}}
\subfloat[TinyImageNet, VGG-16]{\includegraphics[scale=0.25]{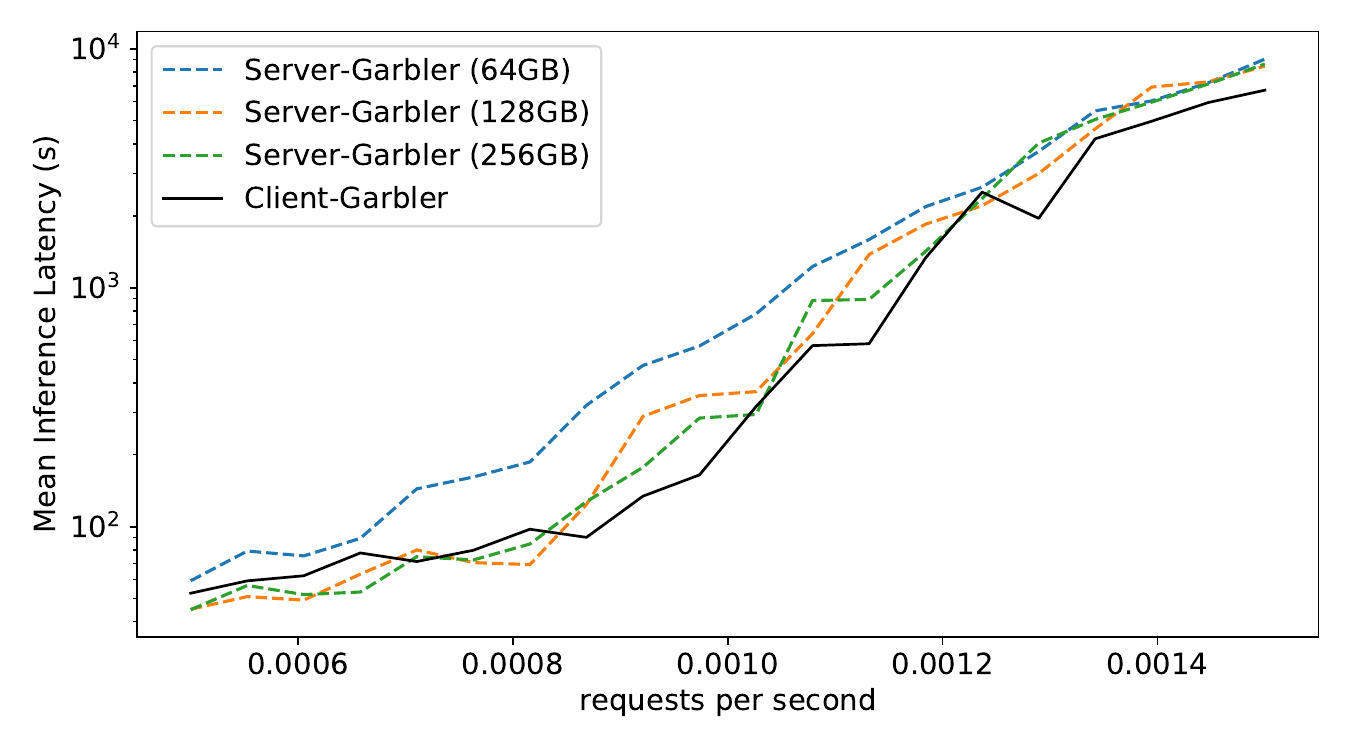}} 
\subfloat[TinyImageNet, ResNet-18]{\includegraphics[scale=0.25]{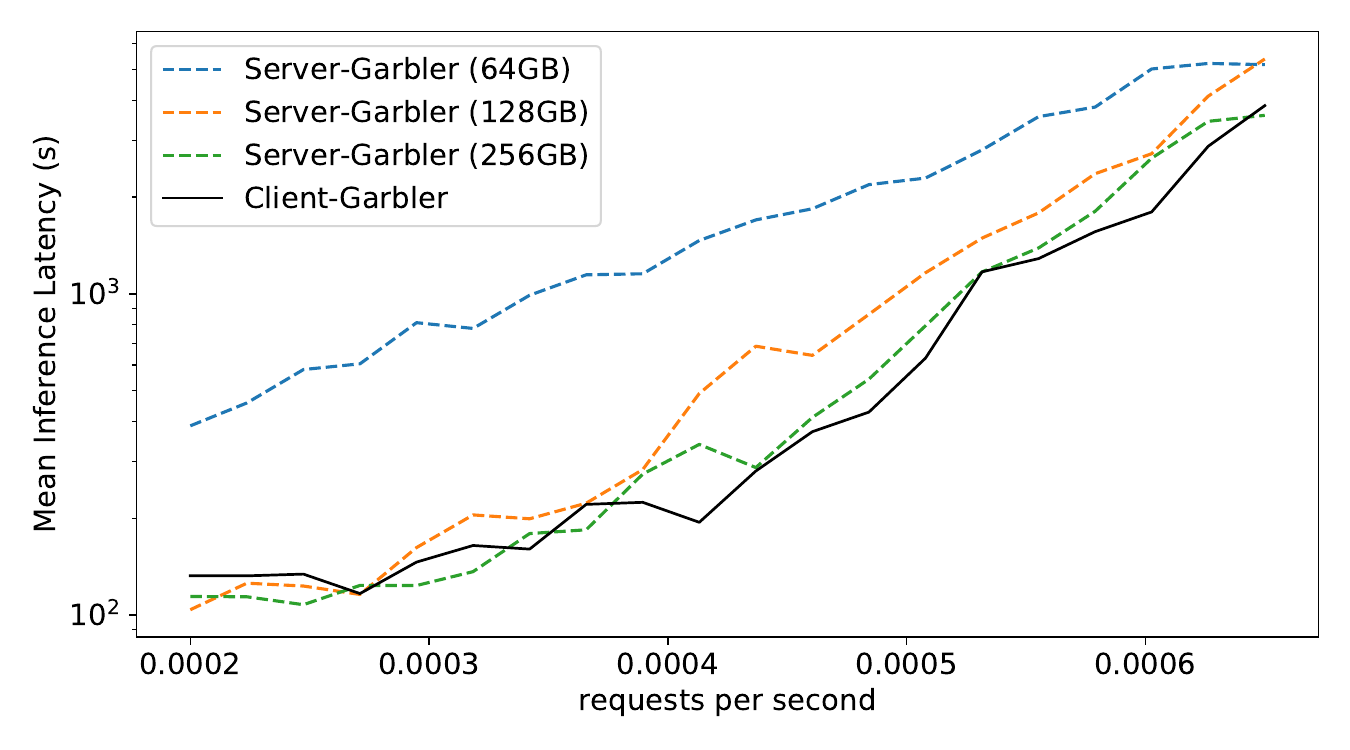}}
\caption{Comparison of the conventional Server-Garbler and our proposed Client-Garbler protocols for single image inference across varying arrival rates for CIFAR-100 (Fig. a-c) and TinyImageNet (Fig. d-f). For the Server-Garbler protocol, the high storage costs of Garbled Circuits prohibits the separation of an offline and online phase even for low to moderate arrival rates. On the other hand, the client-side storage requirements of the Client-Garbler protocol are much lower, mitigating this problem for low to medium arrival rates.
At high arrival rates, HE compute latency dominates overall runtime for both protocols.}
\vspace{-1em}
\label{fig:CompCase2AndCase1}
\end{figure*}

%% file: plots/offline_bottleneck.tex
\begin{figure} [t]
\centering
\includegraphics[scale=0.48]{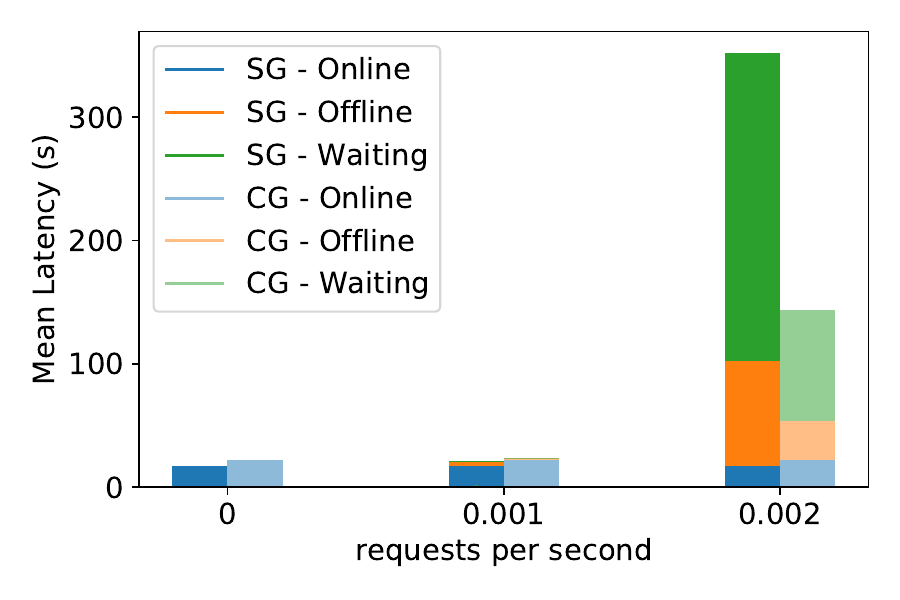}
\vspace{-1em}
\caption{ResNet-18 latency breakdown for the conventional ``Server-Garbler'' (SG) and proposed ``Client-Garbler'' (CG) protocols on CIFAR-100 dataset.}
\label{fig:SgCgWaitingTime}
\end{figure}

%% file: tables/LatencyCost.tex
\begin{table} [t]
\caption{Latency breakdown of Server-Garbler and Client-Garbler protocols for a single inference on CIFAR-100 (C100) and TinyImageNet (Tiny) datasets. The online latency of Client-Garbler is higher than Server-Garbler but the total latency of the former is lower than that of the latter.}
\vspace{.5em}
\label{tab:LatencyCost}
\centering 
\resizebox{0.49\textwidth}{!}{
\begin{tabular}{cccccc} \toprule
\multirow{2}{*}{Dataset} & \multirow{2}{*}{Model} & \multicolumn{2}{c}{Server-Garbler} & \multicolumn{2}{c}{Client-Garbler} \\
\cmidrule(lr{0.5em}){3-4}  
\cmidrule(lr{0.5em}){5-6} 
& & Offline (s) & Online (s) & Offline (s) & Online (s) \\ \toprule
\multirow{3}{*}{C100} & ResNet-32 & 115.2 & 9.4 & 109.1 & 11.9 \\
& VGG-16 & 295.6 & 9.4 & 289.9 & 11.6 \\
& ResNet-18 & 420.8 & 17.2 & 409.6 & 21.8 \\ \midrule
\multirow{3}{*}{Tiny} & ResNet-32 & 401.9 & 39.6 & 377.4 & 49.6 \\
& VGG-16 & 814.7 & 34.2 & 792.2 & 43.4 \\
& ResNet-18 & 1594.0 & 68.5 & 1549.1 & 86.9 \\
\bottomrule
\end{tabular} }
\end{table}

%% file: 09discussion.tex
\section{Discussion and Related work} \label{sec:discussion}

\subsection{Categorization of PI Methods}

\input{tables/optimization_comp}

In this section, we discuss the efficacy of recently proposed PI techniques when considering their full-system effect with non-zero arrival requests rates. 
We categorize previous PI techniques into three categories based on how the FLOPs and ReLU savings affect the GC-cost per ReLU and HE-cost per FLOP depending on arrival rates: (1) low-arrival rate region, (2) moderate-arrival rate region, and (3) high-arrival rate region. 
We now describe the desirable optimization for each of these regions  along with the applicability of PI optimization methods for each category. 

% (see Figure \ref{fig:ArrivalRateRegions})

{\bf Low-arrival rate region:}  As shown in Table \ref{tab:OptimizationComp}, most of the previous PI methods are suitable only for low-arrival rates as they either substitute all the ReLUs with low-degree polynomials (e.g., CryptoNets, SecureML, Polyfit, CryptoDL,  HCNN, Sisyphus, etc.) or reduce the number of non-linear operations by reducing the ReLU count of the network (e.g., DELPHI, CryptoNAS, SAFENets). 
However, in all these optimization strategies, the number of FLOPs (which affects HE costs) is usually disregarded. In fact, some of the ReLU-optimization increase the FLOPs count. For instance, CryptoNAS used neural architecture search and ReLU-balanced layers to find the ReLU-optimized networks while maximizing FLOPs per ReLU; under the assumption that ReLUs dominate PI runtime, this would maximize accuracy per ReLU, a desirable effect.
This works very well for processing inferences in isolation.
However, in a real-world setting where a system is expected to sustain a certain inference throughput, the increase in FLOPs worsens the HE cost, further straining the most costly performance bottleneck of the protocol. 

Another common approach is to replace the PI-unfriendly ReLU activation function with polynomials, which can eliminate GC costs.
It is worth noting that there is usually a significant accuracy penalty for this optimization \cite{garimella2021sisyphus}. Another recent work shows that to achieve reasonable accuracy on the ImageNet-1k dataset \cite{deng2009imagenet}, a polynomial activation function must be of degree $29$ needed\cite{lee2021precise}.
DELPHI and SAFENet~\cite{lou2020safenet} partially substituted some of a network's ReLUs with low-degree polynomials assuming a baseline model is provided as input.
While this will reduce GC costs, the FLOPs count remains the same across all the ReLU-optimized networks, which does not help with HE costs.
On the other hand, CrypTFlow2~\cite{rathee2020cryptflow2} and Circa~\cite{ghodsi2021circa} reduce the GC cost per ReLU computation, rather than replacing them. Thus, both the ReLU and FLOP counts remain the same. MiniONN~\cite{liu2017oblivious}, Gazelle~\cite{juvekar2018gazelle}, and AutoPrivacy~\cite{lou2020autoprivacy} neither reduce ReLU count nor the FLOPs count.
AutoPrivacy does reduce the HE-cost per FLOP by automatically tuning the selection of HE-parameters on a layer-wise basis using the deep reinforcement learning technique.
%layer-wise HE-parameters. 
\input{plots/arrival_rate_regions}

{\bf Moderate-arrival rate region:} 
In this region, we categorize the PI optimization methods with very low GC-cost (networks with only low-degree polynomials or very low ReLU count) and reduce HE-cost (either with reduced HE cost per FLOP or diminished FLOPs count). For instance, Faster-CryptoNets and HEMET have only low-degree polynomial activations and the former reduces the number of HE operations by introducing the sparsity in the network (using pruning and quantization) while the latter reduces the HE cost per FLOPs by reducing the multiplicative depth of the networks. However, the accuracy of Faster CryptoNets degrades substantially on CIFAR-10 and their efficacy on more complex datasets such as TinyImageNet needs to be evaluated. Similarly, HEMET showed the effectiveness on simpler networks such as SqueezeNet \cite{iandola2016squeezenet} and InceptionNet \cite{szegedy2017inception}. 

On the other hand, DeepReDuce (a state-of-the-art in PI) drops the ReLUs by $5\times$ for ResNet-18 on CIFAR-100 and TinyImageNet without loss in accuracy and,  and merges the consecutive linear  (Conv) layers for FLOPs saving with $<1\%$ loss in accuracy. As a result, we believe this a promising direction of reducing both ReLU count (reducing GC costs) and FLOPs count (reducing HE costs).

{\bf High-arrival rate region:}
In this region, we consider the networks with (1) only polynomial activations, reduces HE-cost per FLOP and diminished FLOPs count (for instance Falcon), and (2) very-low ReLU count with reduced GC-cost per ReLU with diminished FLOPs count (e.g., DeepReDuce + Circa, Table 2 in \cite{ghodsi2021circa}). The accuracy of Falcon drops significantly for the CIFAR-10 dataset whereas Circa optimization on DeepReDuce works on complex datasets such as TinyImageNet.  Altogether, with substantially reduced HE-cost along with low GC-cost, PI can be performed with a higher arrival rate. Additionally, we believe that efficient HE compilers \cite{dathathri2020eva,dathathri2019chet,cowan2021porcupine} can be used to further reduce the overall HE-cost and increase the maximal sustainable arrival rate.

In conclusion, low-arrival rate regions appeal to ReLU reduction strategies as Garbled Circuits are the main bottleneck in PI. However, in a real-world scenario which would include moderate to high arrival rates, HE costs move from the offline phase back to the online phase, and thus, it is necessary to optimize for both ReLU \textit{and} FLOPs in order to sustain higher arrival rates.

%% file: tables/optimization_comp.tex
\begin{table*} [t]
\caption{Comparison of the common PI methods from the viewpoint of their efficacy on full-system with non-zero arrival requests. Accuracy comparison are shown on MNIST \cite{deng2012mnist}, CIFAR-10 (C10), CIFAR-100 (C100), and TinyImageNet (Tiny) datasets. Number of $\uparrow$/$\downarrow$ represents the extent of increase/decrease in the quantity compared to the baseline. Most of the previous PI methods performed optimization only for a single PI inference by considering only zero-arrival rate requests.}
\label{tab:OptimizationComp}
\begin{threeparttable}
\centering 
\resizebox{0.99\textwidth}{!}{
\begin{tabular}{c|lcccccccc} \toprule
\multirow{2}{*}{Arrival-rate} & \multirow{2}{*}{ PI method } & \multicolumn{2}{c}{Online cost} & \multicolumn{2}{c}{Offline cost} & \multicolumn{4}{c}{Accuracy} \\
\cmidrule(lr{0.5em}){3-4}  
\cmidrule(lr{0.5em}){5-6} 
\cmidrule(lr{0.5em}){7-10}
& & \#ReLUs &$\frac{GC_{cost}}{\text{ReLU}}$ & \#FLOPs & $\frac{HE_{cost}}{\text{FLOP}}$ & MNIST & C10 & C100 & Tiny \\ \toprule
\multirow{15}{*}{Low}  
& CryptoNets \cite{gilad2016cryptonets} & - & - & $Const$ & $Const$ & $Const$ & - & - & - \\
& SecureML \cite{mohassel2017secureml} & - & - & $Const$ & $Const$ & $\downarrow$ & - & - & - \\
& Polyfit \cite{chabanne2017privacy} & - & - & $Const$ & $Const$ & $\downarrow$ & - & - & - \\ 
& CryptoDL \cite{hesamifard2017cryptodl} & - & - & $Const$ & $Const$ & $Const$ & $Const$ & - & - \\
& Lookup-Table \cite{thaine2019efficient} & - & - & $Const$ & $Const$ & $Const$ & - & - & - \\
& HCNN \cite{badawi2018towards} & - & - & $Const$ & $Const$ & $Const$ & $Const$ & - & - \\
& Sisyphus \cite{garimella2021sisyphus} & - & - & $Const$ & $Const$ & $Const$ & $Const$ & $\downarrow\downarrow$  & $\downarrow\downarrow\downarrow$  \\ 
& MiniONN \cite{liu2017oblivious} & $Const$ & $Const$ & $Const$ & $Const$ & $Const$ & $Const$ & -  & - \\
& GAZELLE \cite{juvekar2018gazelle} & $Const$ & $Const$ & $Const$ & $\downarrow$ & $Const$ & $Const$ & -  & - \\
& DELPHI \cite{mishra2020delphi} & $\downarrow$ & $Const$ & $Const$ & $Const$ & - & $\downarrow$ & $\downarrow$ & - \\ 
& CryptoNAS \cite{ghodsi2020cryptonas}& $\downarrow\downarrow$ & $Const$ & $\uparrow\uparrow$ & $Const$ & - & $Const$ & $Const$ & - \\
& SAFENet \cite{lou2020safenet} & $\downarrow\downarrow$ & $Const$ & $Const$ & $Const$ & - & $\downarrow$ & $\downarrow$ & -\\
& CrypTFlow2 \cite{rathee2020cryptflow2} & $Const$ & $\downarrow$ & $Const$ & $Const$ & - & $Const$ & $Const$ & $Const$\tnote{*} \\
& Circa \cite{ghodsi2021circa} & $Const$ & $\downarrow$ & $Const$ & $Const$ & - & - & $Const$ & $Const$\\
& AutoPrivacy \cite{lou2020autoprivacy} & $Const$ & $Const$ & $Const$ & $\downarrow$ & - & $Const$ & $Const$ & - \\
 \midrule
\multirow{3}{*}{Moderate} 
& F-CryptoNets \cite{chou2018faster} & - & - & $\downarrow$ & $Const$ & $Const$ & $\downarrow\downarrow$ & - & - \\
%& XONN \cite{riazi2019xonn} & $Const$ & - & $Const$ & - & $Const$ & $Const$ & - & - \\
& HEMET \cite{lou2021hemet} & - & - & $Const$ & $\downarrow$ & - & $Const$ & $Const$ & - \\
& DeepReDuce \cite{jha2021deepreduce}& $\downarrow\downarrow\downarrow$ & $Const$ & $\downarrow\downarrow$\tnote{**}& $Const$ & - & - & $\downarrow$&$\downarrow$\\
 \midrule
\multirow{2}{*}{High} 
& Falcon \cite{lou2020falcon} & - & - & $\downarrow$ & $\downarrow$ & $Const$ & $\downarrow$ & - & -  \\ 
& DeepReDuce + Circa & $\downarrow\downarrow\downarrow$ & $\downarrow$ & $\downarrow\downarrow$ & $Const$ & - & - & $\downarrow$ & $\downarrow$ \\
\bottomrule
\end{tabular}}
\begin{tablenotes} \tiny
            \item[*] CrypTFlow2 performed the PI on ImageNet-1k dataset with ResNet-50 and DenseNet-121
            \item[**] DeepReDuce did not reduces the FLOPs for PI but showed the benefits of FLOPs reduction on plaintext (see Table 7 and Table 8 in \cite{jha2021deepreduce})
        \end{tablenotes}
     \end{threeparttable}
\end{table*}

%% file: plots/arrival_rate_regions.tex
% \begin{figure} [t]
% \centering
% \includegraphics[scale=0.5]{figures/arrival_rate_regions.pdf}
% \caption{Arrival-rate regions based on the FLOPs and ReLU count, and HE-cost and GC-cost of PI inference. Networks with higher FLOPs cannot sustain higher arrival rate even when ReLU count and GC-cost per ReLU is less.}
% \label{fig:ArrivalRateRegions}
% \end{figure}

%% file: 10related_work.tex
\section{Related Work}

{\bf Two-party PI methods which do not use HE}
Note that in the aforementioned categorization we do not consider the PI techniques which do not use HE for linear-layer computation. For instance, a few PI methods \cite{riazi2019xonn,Samragh_2021_CVPR} use a Binary neural network (BNN) to reduce GC-cost by a huge margin; however, they showed their efficacy only on MNIST and CIFAR-10. It is well-known that BNN incurs the accuracy drop on complex datasets; even the state-of-the-art BNN suffers accuracy drop on ImageNet \cite{liu2020reactnet}. DeepSecure \cite{rouhani2018deepsecure}, EzPC \cite{chandran2019ezpc}, and Chamelon \cite{riazi2018chameleon} use GC for nonlinear operations and secret-sharing for linear operations. Similarly, CryptGPU \cite{tan2021cryptgpu} and CrypTen \cite{knott2021crypten} use additive secret sharing for linear operations and binary secret sharing for nonlinear operation and rely on trusted-third party (TTP) for generating the beaver triples.

{\bf Private inference with more than two parties} 
There has been a line of research work on three parties  private inference. For instance, FALCON \cite{wagh2021falcon}, CrypTFlow \cite{kumar2020cryptflow}, BLAZE \cite{patra2020blaze}, SecureNN  \cite{wagh2019securenn}, QuantizedNN \cite{dalskov2019secure}, ASTRA \cite{chaudhari2019astra}, ABY3 \cite{mohassel2018aby3}, etc.  Similarly, there has been four party PI such as FLASH \cite{byali2020flash} and Trident \cite{rachuri2019trident}. However, the aforementioned PI methods have weak security model as they sacrificing security by making the assumption of a non-colluding third party.

%% file: 11conclusion.tex
\section{Conclusion}
In this paper, we investigate the end-to-end system characteristics of private inference. 
We find that most recently proposed PI protocols and optimizations focus only on reducing the online inference latency by moving compute-heavy HE cryptographic protocols to an offline phase.
These online-only optimizations work well, 
but only for processing individual inferences, or when assuming an effective arrival rate of zero. 
We find that even for extremely low arrival rates, SOTA PI techniques are unable to endure the storage costs of garbled circuits. Consequently, clients are only able to process very few inferences (at very low arrival rates) at online-only performance before running out of storage and pre-computation before they have to incur both online and offline costs online, which degrades performance by an order of magnitude.
To mitigate the storage issue, we proposed the Client-Garbler protocol to leverage the server's high storage capacity. 
This optimization circumvents the storage bottleneck and enables low-latency inferences at low to moderate arrival rates, reducing  the mean PI latency by $4\times$ compared to the baseline approach.
Finally, we note that recently proposed PI optimizations target online latency and 
ignore full system effects, thus limiting their true benefits.

%% file: ack.tex
\section*{Acknowledgements}
\label{sec:ack}
This work was supported in part by the Applications Driving Architectures (ADA) Research Center, a JUMP Center co-sponsored by SRC and DARPA. This research was
also developed with funding from the Defense Advanced
Research Projects Agency (DARPA),under the Data Protection in Virtual Environments (DPRIVE) program, contract
HR0011-21-9-0003. The views, opinions and/or findings
expressed are those of the author and should not be interpreted as representing the official views or policies of the
Department of Defense or the U.S. Government.